\documentclass{article}
\usepackage[T1]{fontenc}
\usepackage[utf8]{inputenc}
\usepackage{amsmath,amssymb}
\usepackage{graphicx}
\usepackage{lmodern}
\usepackage[unicode]{hyperref}
\usepackage{xcolor}
\usepackage{booktabs}
\usepackage{float}
\usepackage{authblk}

\hypersetup{
pdftitle={Internet Quality Barometer (IQB): A preliminary data-driven evaluation and analysis of the IQB framework},
  hidelinks}
  
\title{Internet Quality Barometer (IQB): A preliminary data-driven evaluation of the IQB framework}
\author{Pavlos Sermpezis\footnote{\href{mailto:pavlos@measurementlab.net}{pavlos@measurementlab.net}}}%
\author{Zeynep Arslan\footnote{\href{mailto:zeynep.arslan@etu.sorbonne-universite.fr}{zeynep.arslan@etu.sorbonne-universite.fr}}}

\affil{Measurement Lab (M-Lab)}

\date{January 2026}

\begin{document}
\maketitle
\section{Introduction}\label{introduction}
The \href{https://www.measurementlab.net/publications/IQB_report_2025.pdf}{\underline{Internet Quality Barometer (IQB) framework}} was designed to transform raw Internet measurement data into actionable insights about Internet quality. 
Specifically, the framework maps raw speed test measurements to network requirements (e.g., throughput, latency), maps these requirements to representative Internet use cases (such as video streaming or web browsing), and finally aggregates performance across use cases into a single IQB score. 
The IQB score is a composite index ranging from 0 to 1, intended to capture overall Internet quality in a way that is both interpretable and comparable across locations.

This mapping relies on a set of parameters that define the IQB framework. 
In the first iteration, these parameters were selected through an extensive consultation process involving more than 60 Internet experts, using a combination of workshops and interviews.
Parameter choices were informed by prior experience, related studies, and general domain knowledge, with the aim of approximating how different network performance levels translate into perceived user experience. 
However, during this initial phase it was not possible to empirically evaluate the resulting IQB scores, or to assess how sensitive they were to specific parameter choices. 
This was because the focus of the first phase was on conceptual framework design, rather than on developing the tools required to compute IQB scores at scale.

In the second phase of the project, we implemented the IQB framework in practice by developing an \href{https://github.com/m-lab/iqb}{\underline{open-source IQB library}} (\href{https://github.com/m-lab/iqb}{\underline{https://github.com/m-lab/iqb}}) and a \href{https://iqb.mlab-staging.measurementlab.net/}{\underline{prototype web application}} (\href{https://iqb.mlab-staging.measurementlab.net/}{\underline{https://iqb.mlab-staging.measurementlab.net/}}).
These tools enabled us to compute IQB scores at scale, including global estimates aggregated at the level of countries, regions, cities, and, where data permitted, individual Internet service providers (ISPs). 
With these computed scores, we can now begin to assess whether the initial design choices achieve their intended objectives, and to what extent the resulting patterns align with common intuition and existing
understanding of Internet connectivity across different parts of the
world.

The primary objective of this report is to conduct a preliminary sensitivity analysis of the IQB framework. 
We aim to develop an initial understanding of how different parameter choices affect the resulting IQB scores, to identify which parameters the framework is most sensitive to, and to highlight cases that may lead to outliers or potentially misleading results.
In addition, we explore the impact of data availability on the robustness of the scores, for example by comparing regions with sparse measurements to those with extensive data coverage.

In the remainder of this report, we first outline the research questions guiding our analysis, and then present the results of our preliminary evaluation.

\section{Related Work and Background }\label{related-work-and-background}

The challenge of measuring Internet quality at scale has attracted sustained attention from researchers, regulators, and industry. 
Early efforts focused almost exclusively on throughput (raw download and upload speeds) as the primary indicator of broadband performance. 
This framing shaped the dominant measurement platforms and remains embedded in how most stakeholders think about Internet quality today. 
However, a growing body of work has demonstrated that throughput alone fails to capture the multi-dimensional nature of user experience, particularly for latency-sensitive or interactive applications.

The two most widely used sources of Internet performance data are Ookla's Speedtest platform and Measurement Lab's (M-Lab) Network Diagnostic Tool (NDT).
Ookla's Speedtest Global Index ranks countries by median download speed based on user-initiated tests collected through its applications, using a "Precision Threshold" methodology that requires the 95\% confidence interval of a country's median to fall within ±5\% before inclusion \cite{ookla}.
M-Lab's NDT is an open-source diagnostic tool that measures throughput, latency, and packet loss using a single TCP connection; its data is publicly available and has been used extensively in academic research \cite{clark2021, gill2022}.

Both platforms rely on crowdsourced, user-initiated tests, which introduces well-documented biases. Users disproportionately run speed tests when experiencing connectivity problems, which can skew lower percentiles of the measurement distribution downward \cite{gill2022, paul2022}. M-Lab's NDT additionally uses a single TCP connection, which can underreport capacity compared to Ookla's multi-connection approach \cite{macmillan2023comparative}.
Furthermore, crowdsourced tests are not uniformly distributed across populations: geographic coverage is uneven, and participation correlates with income, urbanization, and device type, meaning the measurement population may not be representative of the broader user base \cite{paul2022characterizing}. 
M-Lab has also noted server-side biases, as its servers are placed at transit ISPs, potentially producing paths longer than those users experience when accessing popular content \cite{gill2022}.

These biases have practical consequences for any framework that derives aggregate scores from crowdsourced data. 
They underscore the importance of the percentile choice examined in this report: higher percentiles (e.g., the 95th) may partially mitigate the effect of problem-motivated testing, while lower percentiles are more susceptible to this bias but may better reflect typical user experience.

\textbf{Composite ICT and Internet Quality Indices}

The construction of composite indices to summarize multi-dimensional phenomena is a well-established practice across domains, from economics (the Human Development Index) to public health. 
The OECD and UNECE have published detailed guidelines for composite index construction, identifying ten standard steps including conceptual framework development, data selection, normalization, weighting, and aggregation, followed by sensitivity analysis and validation \cite{un2019,oecd2008}.

In the ICT domain, the most prominent composite index is the International Telecommunication Union's ICT Development Index (IDI), which combines 11 indicators across three sub-indices (access, use, and skills) into a single country-level score \cite{itu}. 
The IDI has been widely used but also criticized: Gerpott and Ahmadi \cite{gerpott2015} argued that the ITU's weighting scheme overemphasizes usage and skills relative to accessibility, efficiency, and quality, and proposed an alternative index that incorporates broadband speed and call reliability indicators. 
Their work demonstrated that different weighting choices can produce substantially different country rankings - a finding directly relevant to the IQB framework's own sensitivity to use-case weights.

Other indices relevant to the IQB context include the Alliance for Affordable Internet (A4AI) Affordability Drivers Index, which incorporates policy and pricing dimensions alongside infrastructure, and cable.co.uk's broadband speed rankings, which aggregate Ookla data into country-level comparisons. 
However, none of these indices adopt the use-case-driven approach that is central to IQB: they measure Internet characteristics (speed, price, penetration) rather than the extent to which connectivity supports specific user activities.

\textbf{Quality of Experience (QoE) Frameworks}

The QoE research community has developed a rich theoretical and empirical foundation for understanding how network-level metrics translate into user-perceived quality. 
The ITU defines QoE as the "overall acceptability of an application or service, as perceived subjectively by the end-user" {[}ITU-T Rec. P.10/G.100, 2016{]}, and the COST Action QUALINET further refined this to emphasize the role of user expectations and context \cite{moller2014, qualinet2013}.

A central challenge in QoE research is the mapping from objective, measurable Quality of Service (QoS) parameters---throughput, latency, jitter, packet loss---to subjective user satisfaction.
This mapping is typically nonlinear and application-dependent. 
For video streaming, buffering events and resolution switches are the dominant QoE factors; for web browsing, page load time drives user satisfaction; for gaming, latency and its variability are critical \cite{duanmu2017, yang2018}.
These application-specific relationships have been modeled extensively, often using Mean Opinion Score (MOS) scales validated through subjective user studies.

The IQB framework draws on this tradition by decomposing Internet quality into use cases with distinct network requirements, each defined by thresholds for throughput, latency, and packet loss. 
This approach is conceptually aligned with use-case-specific QoE modeling, but differs in an important respect: IQB uses expert-defined thresholds and weights rather than empirically calibrated QoE models derived from user studies.
The Cranor et al. \cite{cranor2022} study on broadband consumer labels, which informed IQB's choice of six use cases, found that consumers care about performance in relation to specific activities (streaming, gaming, video calls) and want to understand service quality in those terms rather than through raw technical metrics.

Several recent frameworks have attempted to bridge the gap between per-application QoE assessment and aggregate network quality scoring.
Wang et al.\cite{wang2023} proposed a general QoE assessment framework for applications and services using a deconstruction-and-aggregation approach, where individual application QoE scores are combined into an overall assessment.
Similarly, the unified QoE scoring framework of Xu et al. \cite{xu2019} classifies network traffic by type and applies type-specific nonlinear QoS-to-QoE mappings. 
These works share IQB's goal of producing aggregate quality assessments from heterogeneous application requirements, though they typically operate at the individual connection level rather than at the regional or national scale that IQB targets.

\textbf{Percentile Aggregation in Network Measurement}

The choice of which summary statistic to use when aggregating network measurements is a recurring methodological question in broadband measurement research.
The FCC's MBA program transitioned from reporting mean speeds to median speeds in 2016, noting that the two were typically within 5\% of each other but that medians are more robust to outliers \cite{fcc2016}. 
Ookla's Speedtest Global Index uses median download speed as its primary ranking metric.
IQB's initial choice of the 95th percentile as its aggregation threshold represents a deliberate decision to approximate infrastructure capacity rather than typical user experience. 
This choice differs from the median or 75th percentile used by most other measurement programs, and the sensitivity analysis presented in this report examines its implications.
The broader literature on composite index construction suggests that the choice of aggregation method can substantially affect rankings and should be subjected to systematic robustness testing---a recommendation that directly motivates the analysis in Section 3.

\section{Research questions}\label{research-questions}

This preliminary analysis focuses on assessing whether the design choices made in the first iteration of the Internet Quality Barometer (IQB) framework lead to results that are robust, interpretable, and aligned with intuitive expectations about Internet quality. 
Given the exploratory nature of this phase, our goal is not to definitively validate the framework, but rather to identify sensitivities, potential biases, and open questions that should inform future refinements. 
To this end, we structured our analysis around a set of research questions:

\textbf{Threshold/Percentile sensitivity and validation:} IQB scores are
derived from aggregate statistics computed over large collections of
speed test measurements.
A key design decision concerns which percentile (or, ``threshold'') of the measurement distribution should be used to represent network performance (e.g., median performance versus tail performance).

In this analysis, we examine the impact of using different percentiles (e.g., 50th, 75th, or 90th percentiles) on the resulting IQB scores.
The objective is to justify the initial choice of percentile by assessing how sensitive country-level rankings are to this parameter.
We ask whether rankings change substantially when alternative percentiles are used, and whether certain regions or countries are disproportionately affected by this choice.

The purpose of this analysis is to evaluate the robustness of IQB scores to changes in these reference values.
In particular, we investigate whether relative comparisons remain stable when thresholds are varied within plausible ranges.
The key research question is whether changes in reference values significantly affect relative rankings across countries, regions, or providers, or whether the framework preserves broad ordering even under different parameter choices.

\textbf{Use-case validation:} The purpose of this analysis is to assess whether the IQB framework meaningfully differentiates between use cases and produces plausible, intuitive results. 
We examine which countries rank highest for different activities. 
This analysis provides an initial check on whether the use case abstractions capture relevant aspects of real-world Internet usage.

\textbf{Measurement uncertainty quantification:} Finally, IQB scores are computed from empirical measurement data whose volume and quality vary substantially across regions. 
As a result, scores derived from sparse data may be subject to higher uncertainty than those based on large sample sizes.
We investigate how sample size affects scores. 
A key question is whether minimum sample size thresholds can be identified below which IQB scores should be interpreted with caution.

\section{Results}\label{results}

\subsection{\texorpdfstring{How aggregation percentiles affect IQB
scores?}{How aggregation percentiles affect IQB scores? }}\label{how-aggregation-percentiles-affect-iqb-scores}

The first iteration of the IQB framework recommended using the 95th percentile as the aggregation threshold for speed test measurements.
Under this approach, network performance metrics are first aggregated into a distribution for a given geographic region, after which the 95th percentile---representing the best observed performance---is selected and compared against the minimum quality thresholds defined for each use case.

The rationale for choosing the 95th percentile is to approximate the underlying infrastructure capacity of a region, rather than its typical or average performance.
This choice emphasizes what the network is capable of delivering under favorable conditions, abstracting away from transient congestion or localized issues.
However, to obtain a more representative view of Internet connectivity as experienced by the broader population, lower percentiles (such as the 75th or 50th) may be more appropriate.

Interpreting lower-percentile results requires particular care due to known measurement biases in crowdsourced speed test data. 
In particular, users tend to initiate speed tests more frequently when experiencing connectivity problems, which can disproportionately affect lower percentiles.
These trade-offs highlight the importance of explicitly aligning percentile choices with the intended analytical or policy-making objective.

\subsubsection{The global IQB map}\label{the-global-iqb-map}

The prototype web application visualizes IQB scores at the country level using a global map representation. 
Figure 1 shows this map under two different aggregation choices: IQB scores computed using the 95th percentile (top) and the 75th percentile (bottom).

In the former case, nearly all countries exhibit very high IQB scores (shown in blue). 
This is expected, as the 95th percentile reflects the best-performing segment of connectivity within each country. 
In contrast, the map based on the 75th percentile reveals substantially lower IQB values for several countries, particularly in parts of Africa and Central Asia. 
This percentile captures a broader portion of the population and thus reflects more typical, rather than peak, Internet performance.

These visual differences motivated a deeper investigation into how aggregation thresholds influence the insights derived from IQB scores.
In particular, they highlight the extent to which percentile choice can affect the apparent global distribution of Internet quality and the conclusions drawn from comparative analyses.

\begin{figure}[H]
  \centering
  \includegraphics[width=\linewidth]{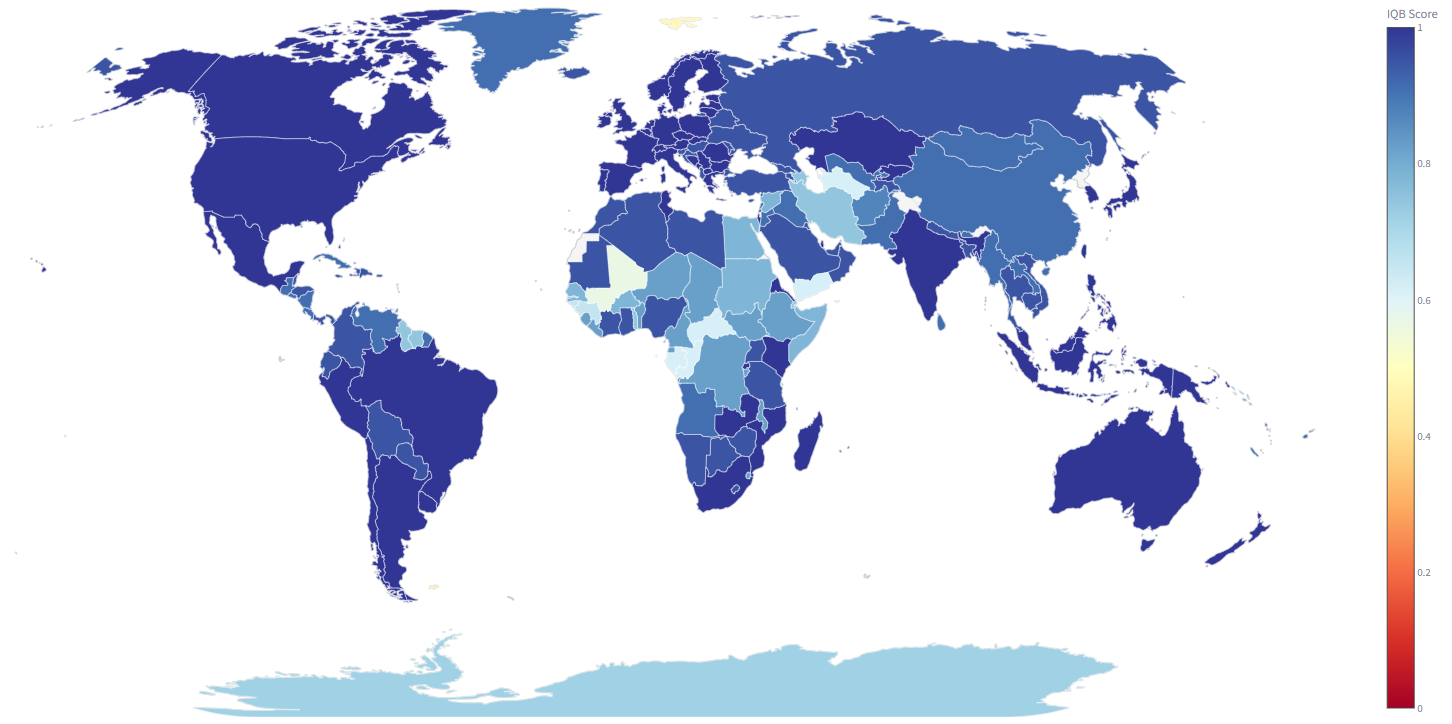}
  \includegraphics[width=\linewidth]{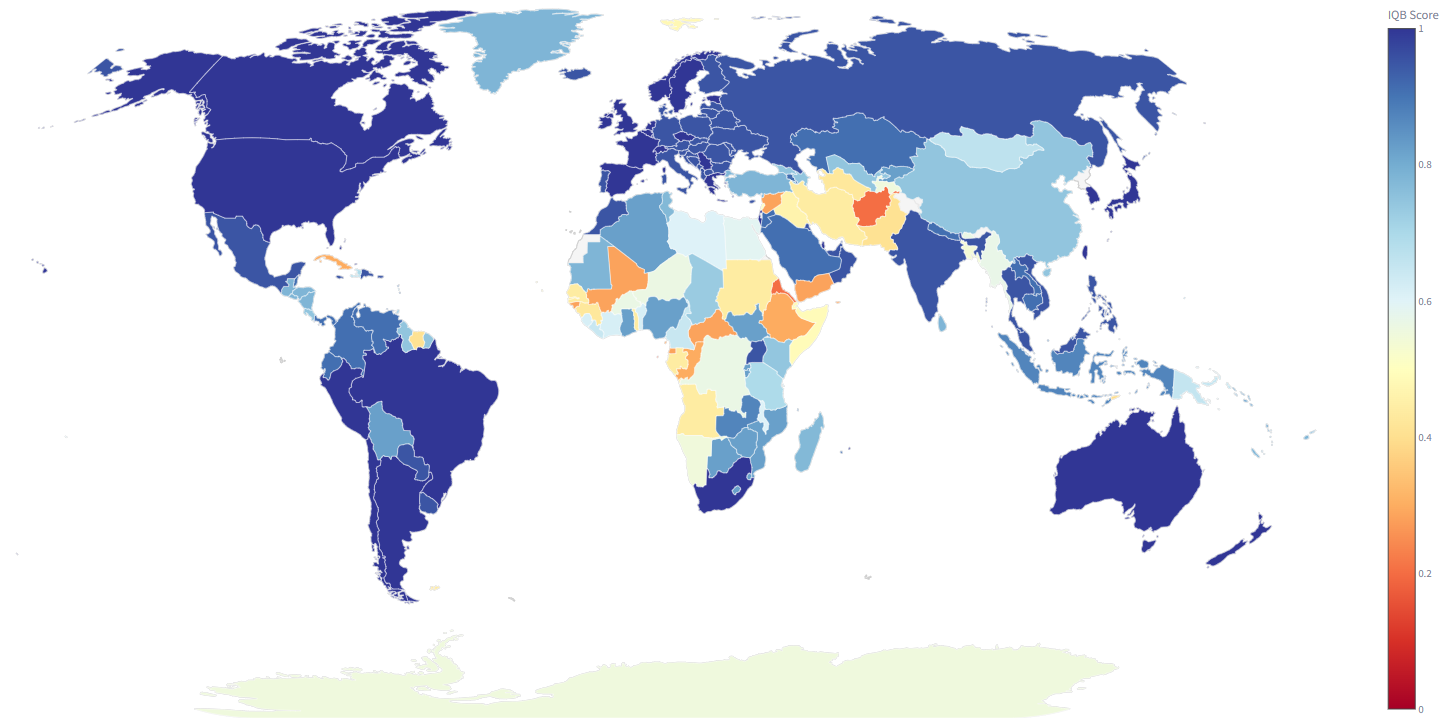}
  \caption{The global Internet Quality Barometer map. IQB scores calculated using the 95th (top map) and 75th percentiles (bottom map).}
  \label{fig:1}
\end{figure}

\subsubsection{Aggregation percentiles vs IQB scores: an example.}\label{aggregation-percentiles-vs-iqb-scores-an-example.}

We compute IQB scores using different aggregation percentiles to examine how Internet quality varies across different segments of the population.
Figure 2 shows IQB scores (y-axis) as a function of the chosen percentile (x-axis) for two countries: the United States (left) and Zambia (right).

As expected, higher percentiles correspond to higher IQB scores, reflecting better performance among the more well-connected users.
However, the rate at which scores increase differs substantially between the two countries. 
In the United States, IQB scores reach the maximum value of 1 already at the 75th percentile, indicating that a large fraction of users experience near-optimal Internet quality. 
In contrast, Zambia reaches a perfect IQB score only at the 90th percentile, suggesting that high-quality connectivity is limited to a smaller subset of the population.

\begin{figure}[H]
  \centering
  \begin{minipage}[b]{0.49\linewidth}
    \centering
    \includegraphics[width=\linewidth]{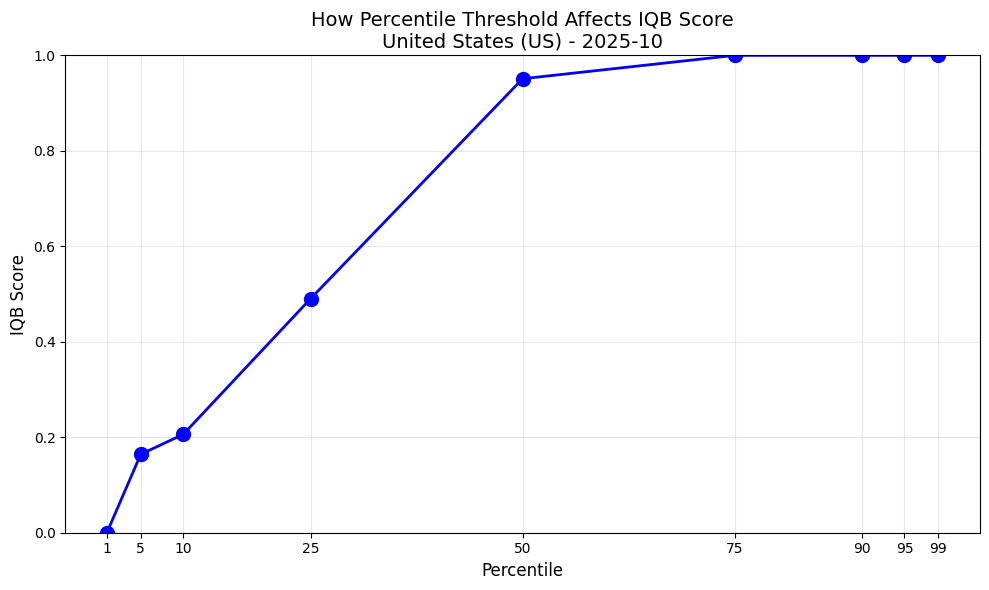}
  \end{minipage}\hfill
  \begin{minipage}[b]{0.49\linewidth}
    \centering
    \includegraphics[width=\linewidth]{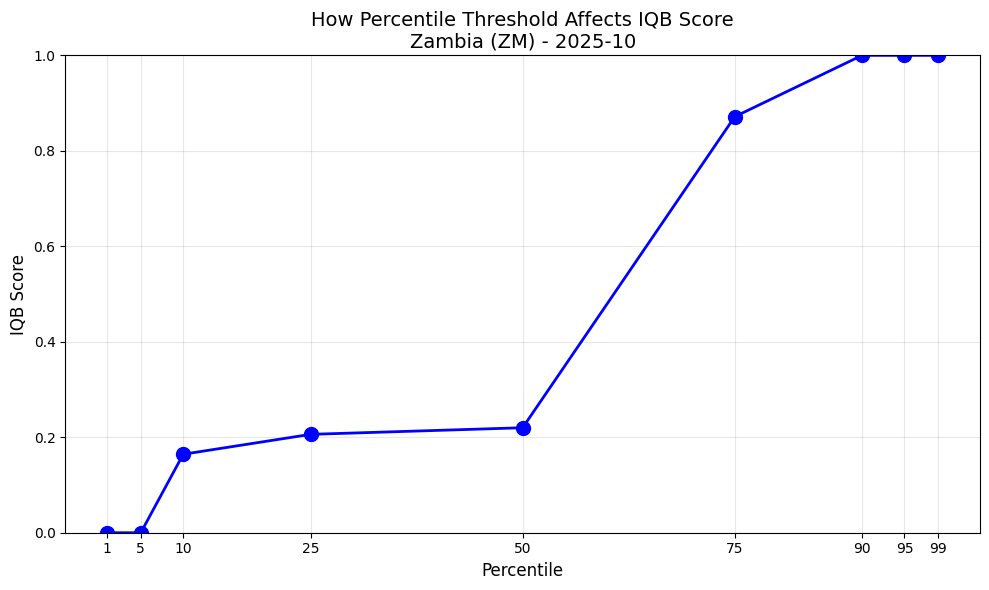}
  \end{minipage}
  \caption{Aggregation percentiles vs IQB scores: IQB scores (y-axis) calculated with different percentiles (x-axis) for the countries of the United States (left) and Zambia (right)}
  \label{fig:2}
\end{figure}

These IQB score vs. percentile plots provide a useful way to compare Internet quality across different population segments within the same region, offering a potential lens on digital inequality. 
By contrasting performance for lower versus higher percentiles, the framework can highlight disparities between poorly and well-connected users that are not visible in single-point summaries.

To further illustrate this point, Figure 3 compares IQB scores computed across multiple percentiles for two U.S. states: New York and Montana.
This comparison allows us to explore intra-country differences in Internet quality distribution and to assess whether similar aggregate scores may mask substantial differences in how connectivity is experienced across populations.

\begin{figure}[H]
  \centering
  \includegraphics[width=\linewidth]{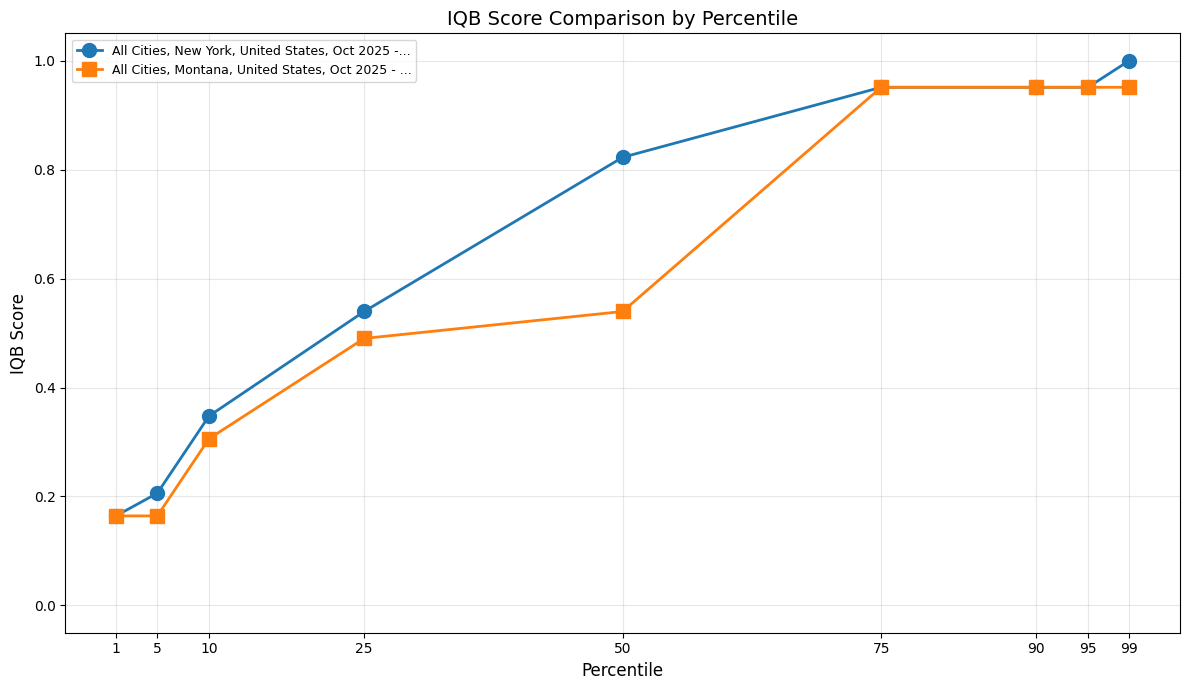}
  \caption{Aggregation percentiles vs IQB scores: Comparison of IQB scores for New York (blue line) and Montana (orange line) calculated with different percentiles (x-axis)}
  \label{fig:3}
\end{figure}

Similar plots can be calculated for any country or city (and their
comparisons) with our tools
(\href{https://colab.research.google.com/drive/1YLYpCtrW1s2px29KdzJGA0NGPVqieOFL?usp=sharing}{\underline{Google
Colab}}).

\subsubsection{Aggregation percentiles vs IQB scores: per country distribution.}\label{aggregation-percentiles-vs-iqb-scores-per-country-distribution.}

We next extend the analysis to all countries worldwide and compute IQB scores using different percentile values.
Figure 4 shows the resulting score distributions.

When using the 75th percentile (rightmost plots), more than 30\% of countries achieve near-perfect IQB scores. 
This observation is consistent with the patterns seen in Figure 1 and suggests that higher percentiles tend to emphasize peak or near-peak network performance. 
As such, this percentile choice may be more representative of a country's underlying infrastructure capacity rather than typical user experience.

In contrast, when using the 50th percentile (middle plots), the distribution of IQB scores is more dispersed. 
This increased spread results in greater differentiation between countries, which may make this percentile more suitable for comparative analyses or ranking exercises.
However, it also reflects median performance, which may underrepresent best-case capabilities in some contexts.

Despite these insights, the results remain inconclusive with respect to identifying a single ``best'' percentile choice. 
A more in-depth investigation is required, potentially incorporating expert judgment and comparisons with external datasets, to guide the selection of percentile values tailored to specific policy-making objectives.

\begin{figure}[H]
  \centering
  \includegraphics[width=\linewidth]{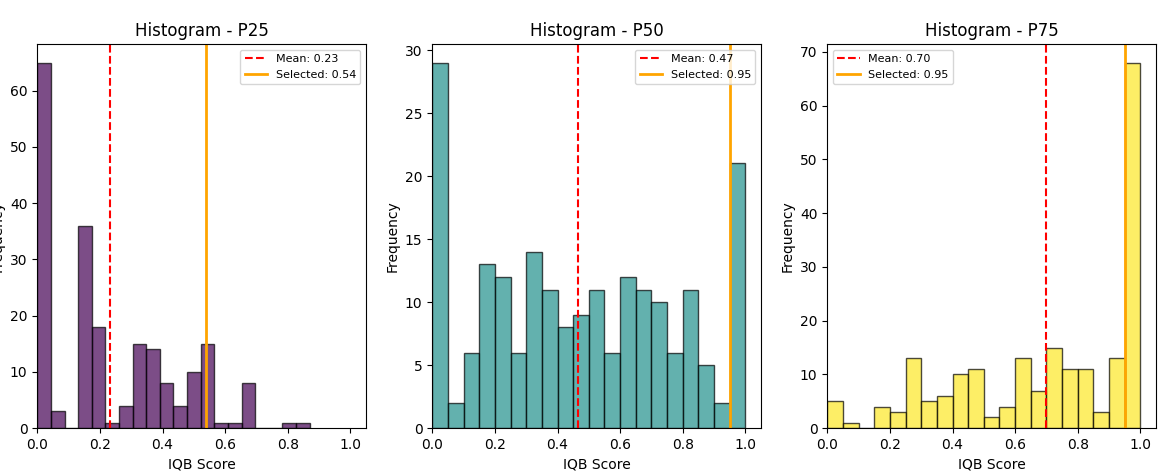}
  \includegraphics[width=\linewidth]{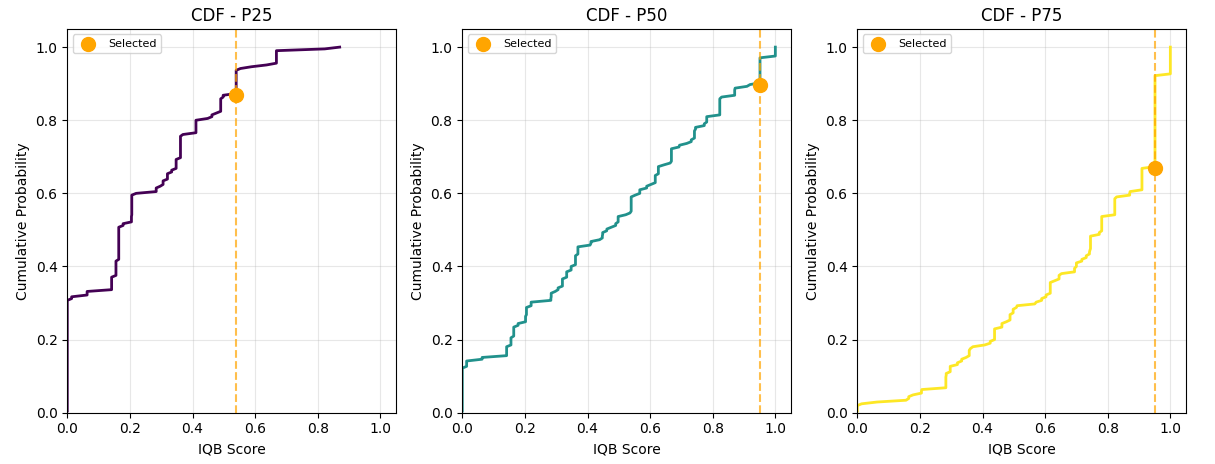}
  \caption{Distributions of the IQB scores for all countries--presented as histograms (top) or CDFs (bottom)--calculated under three different percentiles: 25th (left), 50th (middle), and 75th (right). The vertical lines correspond to the scores of the US.}
  \label{fig:4}
\end{figure}

\subsubsection{Data aggregation: different percentiles, similar insights?}\label{data-aggregation-different-percentiles-similar-insights}

While different percentile choices can be appropriate for addressing different policy questions, the IQB framework should, in principle, yield qualitatively consistent results across reasonable percentile selections. 
To assess this, Figure 5 compares country-level IQB scores computed using the 75th percentile (y-axis) and the 50th percentile (x-axis), with each dot representing a country.

The results show a strong positive correlation between the two sets of scores (Pearson's correlation coefficient = 0.8), indicating that overall patterns and relative performance are largely preserved. 
This provides encouraging evidence for the robustness of the IQB framework with respect to percentile choice.

At the same time, the scores are not perfectly aligned across percentiles.
For some countries, relative ordering changes depending on whether median or higher-end performance is emphasized. 
This suggests that percentile selection can meaningfully affect comparative assessments, particularly when the analysis is used for ranking, prioritization, or resource allocation across regions.

These findings highlight the importance of interpreting IQB results in light of the underlying percentile choice. 
For more robust evaluations, especially in comparative contexts, analyses should consider multiple percentiles and synthesize their results rather than relying on a single summary statistic.

\begin{figure}[H]
  \centering
  \includegraphics[width=0.75\linewidth]{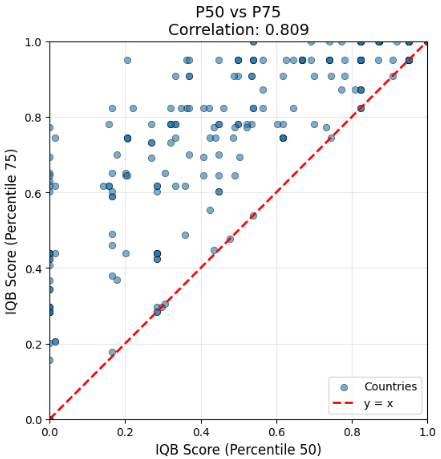}
  \caption{IQB scores for all countries (dots) calculated under the 75th (y-axis) and 50th (x-axis) percentile.}
  \label{fig:5}
\end{figure}

\subsection{{IQB scores per use case }}\label{iqb-scores-per-use-case}
\subsubsection{Different use cases, different insights}\label{different-use-cases-different-insights}

Different use cases contribute differently to the overall IQB score, reflecting their distinct network performance requirements. 
Figure 6 illustrates this by showing country-level scores for two representative use cases: (i) web browsing, which has relatively modest network demands (left plot), and (ii) online gaming, which imposes the most stringent requirements in the framework (right plot).

As expected, web browsing scores are generally high, with more than 50\% of countries achieving a perfect score of 1. 
This indicates that basic Internet functionality is widely supported across regions. 
In contrast, gaming scores exhibit a much broader and nearly uniform distribution.
Only around 10\% of countries achieve a perfect score for this use case, while nearly half of the countries have scores below 0.5, highlighting substantial limitations in latency-sensitive performance worldwide.

These results illustrate how the IQB framework differentiates between use cases with varying performance demands and helps surface aspects of Internet quality that are not captured by aggregate metrics alone.

\begin{figure}[H]
  \centering
  \includegraphics[width=0.48\linewidth]{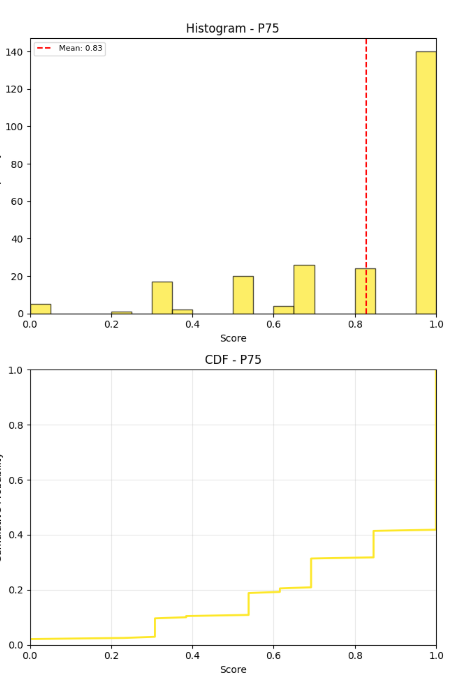}
  \includegraphics[width=0.48\linewidth]{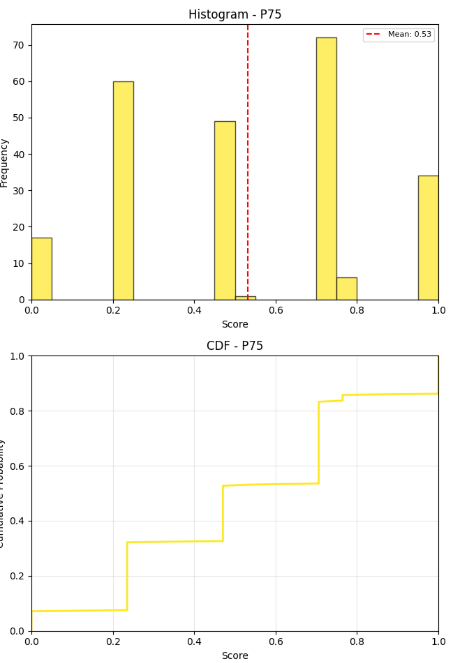}
  \caption{Distributions of the \emph{use case scores} for the use cases of Web browsing (left) and Gaming (right) for all countries--presented as histograms (top) or CDFs (bottom)--calculated with the 75th percentile.}
  \label{fig:6}
\end{figure}

\subsubsection{Country comparison per use case: an example}\label{country-comparison-per-use-case-an-example}

Figure 7 compares use case--specific IQB scores, calculated using the 50th percentile, for the United States (left) and Brazil (right) across six representative Internet use cases.

Both countries exhibit strong performance for less demanding activities such as web browsing, video streaming, and audio streaming, where median performance is sufficient to meet use case requirements.
In these categories, scores are relatively high and comparable between the two countries, suggesting broadly similar support for everyday Internet usage at the median level.

More pronounced differences emerge for use cases with stricter network requirements. 
In particular, gaming shows lower scores in both countries, reflecting the sensitivity of this use case to latency.
The United States achieves a moderately lower score than Brazil, indicating better median performance for latency-sensitive applications.
Differences are also visible for video conferencing and online backup, where the United States exhibits more balanced performance across use cases, while Brazil shows slightly greater variability.

Overall, this comparison illustrates how the IQB framework can be used to move beyond aggregate country-level scores and highlight use case--specific strengths and limitations. 
It also demonstrates how similar overall Internet quality can translate into different experiences depending on the activity considered, even when evaluated at the same aggregation percentile.

\begin{figure}[H]
  \centering
  \includegraphics[width=0.8\linewidth]{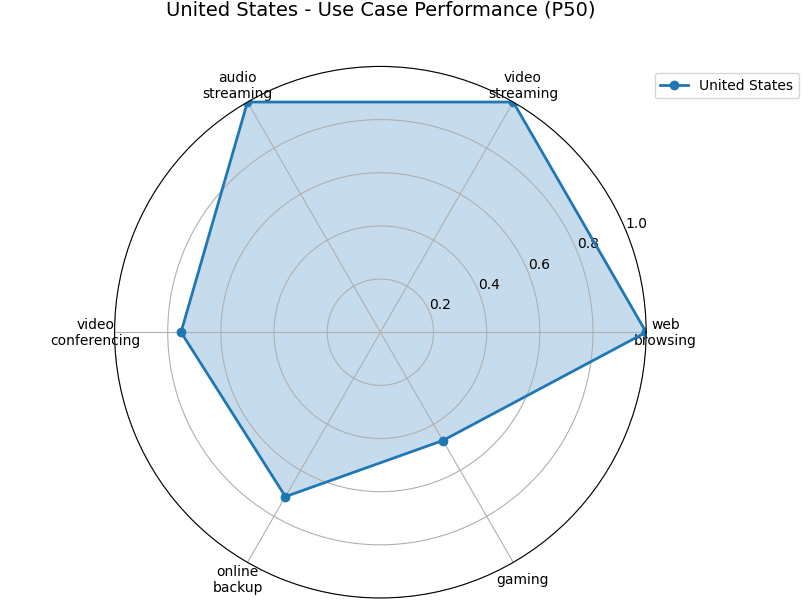}
  \includegraphics[width=0.8\linewidth]{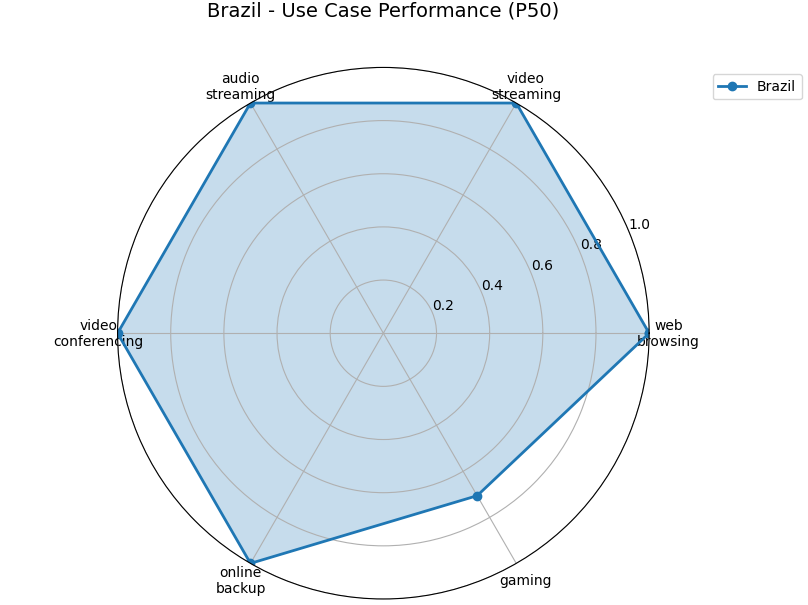}
  \caption{Comparison of the \emph{use case scores}--calculated with the 50th percentile--for all the six use cases for the US (left) vs Brazil (right).}
  \label{fig:7}
\end{figure}

\subsubsection{Aggregation percentiles vs Use case scores: global trends}\label{aggregation-percentiles-vs-use-case-scores-global-trends}

Figure 8 presents average use case scores across all countries for the six use cases, computed using different aggregation percentiles. 
Scores are color-coded from low (red) to high (green), highlighting how both use case requirements and percentile choice influence the resulting IQB values. 
Overall, the figure illustrates the combined impact of percentile choice and use case requirements on IQB outcomes.

Across all use cases, scores increase with higher percentiles; however, the rate of improvement differs substantially between use cases. 
Less demanding activities, such as web browsing, reach high average scores already at intermediate percentiles (around the 75th percentile),
indicating that these services are widely supported across countries. 
In contrast, more demanding use cases (e.g., video conferencing or gaming) exhibit significantly lower scores at lower and median percentiles.

\begin{figure}[H]
  \centering
  \includegraphics[width=1\linewidth]{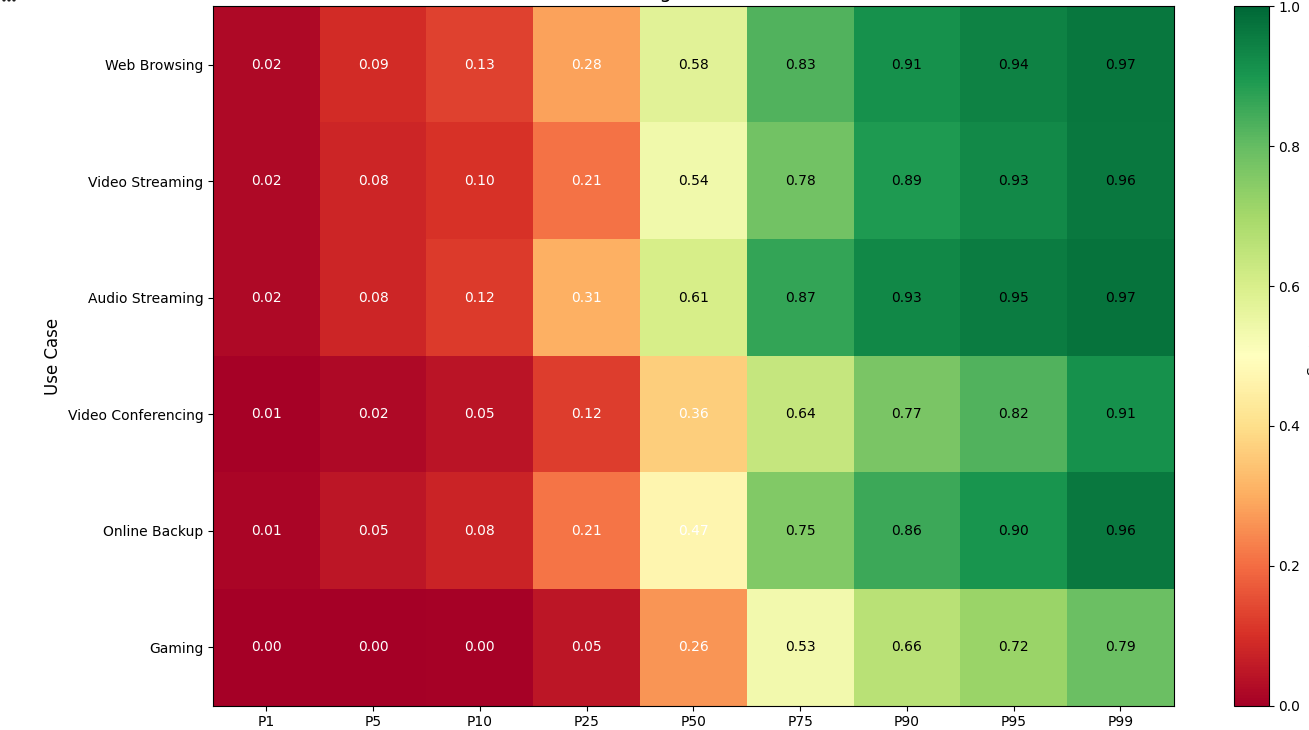}
  \caption{Average use case scores across all countries (high-green, low-red) for different use cases (y-axis) calculated with different percentile values (x-axis)}
  \label{fig:8}
\end{figure}

\subsection{Robustness of IQB calculations to number of samples}\label{robustness-of-iqb-calculations-to-number-of-samples}

Figure 9 shows country-level IQB scores (y-axis) as a function of the number of available speed test measurements (x-axis), computed using the 75th percentile. 
Each point represents a country, and the dashed line indicates a smoothed trend.

Overall, the correlation between IQB score and sample size is low (Pearson coefficient = 0.15). 
This weak correlation is a positive result, as it suggests that IQB scores are not strongly driven by the volume of available measurements alone.
In other words, countries with large numbers of measurements do not systematically receive higher scores, indicating a degree of robustness of the framework to uneven data availability across regions.

At the same time, the figure does not reveal a clear threshold for the minimum number of samples required to obtain stable or reliable IQB scores. 
Countries with relatively few measurements can exhibit both high and low scores, and substantial variability remains even as sample sizes increase.
Based on these observations, we believe that at least several hundreds of measurements are likely necessary to support meaningful IQB estimates, although this should be viewed as a tentative guideline rather than a strict requirement.

Finally, while this high-level analysis provides initial reassurance, it is not sufficient to draw definitive conclusions about measurement uncertainty. 
A more detailed variation analysis and sensitivity assessment---examining how IQB scores fluctuate as a function of sample size, percentile choice, and other parameters---is needed to more rigorously quantify confidence and establish data sufficiency criteria.

\begin{figure}[H]
  \centering
  \includegraphics[width=0.8\linewidth]{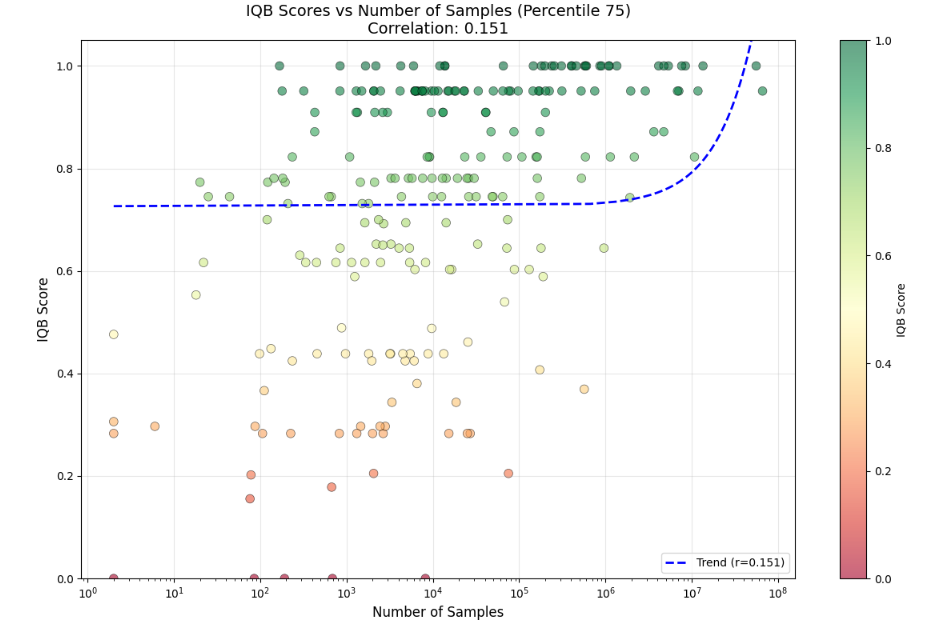}
  \caption{IQB scores (y-axis) vs number of samples (x-axis) per country}
  \label{fig:9}
\end{figure}

\subsection{A first step to longitudinal analysis}\label{a-first-step-to-longitudinal-analysis}

Figure 10 compares the IQB scores per country between Oct 2024 and Oct 2025. 
While many countries maintain similar Internet quality levels (values near the diagonal in the scatter plot) , there are notable differences in some cases.

The CDF of the difference scores further illustrates these patterns, showing that the majority of countries have differences near zero, approximately 30\% of countries have experienced improvements in Internet quality, whereas around 10\% of countries show a decline. 
A few cases exhibit relatively large increases or decreases in scores. 
It is important to interpret these extreme changes cautiously, as differences greater than 0.2 may be suspicious and could be influenced by a low number of samples.
Investigating these outliers in future analyses will help clarify whether they represent genuine shifts in Internet quality or data artifacts.

Nevertheless, this type of analysis can provide valuable longitudinal insights by enabling comparisons of Internet quality across different countries over time. 
Tracking changes in IQB scores across multiple periods allows us to identify trends, highlight countries with consistent improvements or declines, and flag potential outliers for further investigation. 
Over time, such analyses can help policymakers and researchers understand the dynamics of Internet performance and target interventions where they are most needed.

\begin{figure}[H]
  \centering
  \includegraphics[width=0.75\linewidth]{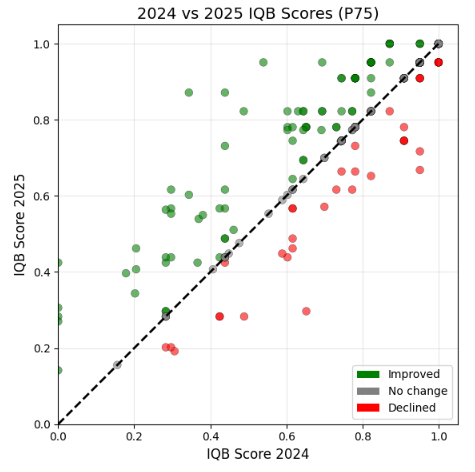}
  \includegraphics[width=0.75\linewidth]{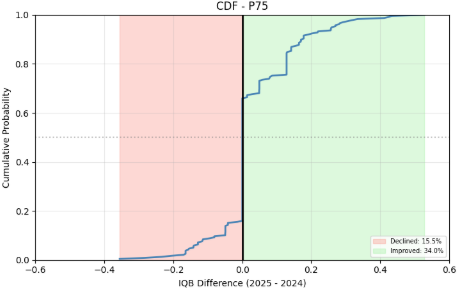}
  \caption{Comparison of IQB scores per country between Oct 2024 and Oct 2025. Left: scatter plot of scores for the different years; Right: distribution (CDF) of the difference score\_2025 - score\_2024 across all countries.}
  \label{fig:10}
\end{figure}

\section{Future Directions}\label{future-directions}

The preliminary analysis presented in this report points to several directions for future work.
In the following we discuss three research questions that we consider of particular significance. 
Nevertheless, these research directions are intended to be indicative rather than exhaustive; many additional aspects warrant investigation.

\textbf{IQB vs development context:} The sensitivity analysis should be extended to examine whether percentile sensitivity varies systematically by development context, with countries exhibiting heterogeneous infrastructure potentially showing greater score volatility across threshold choices; related to this, identifying the percentile values at which countries cross minimum or high-quality classification boundaries, and establishing minimum sample-size guidelines through bootstrap resampling, would strengthen the framework's accountability properties.

\textbf{Context-specific use case ``profiles''}: The assumption of uniform use-case weighting warrants investigation: context-specific profiles that de-prioritise use cases with limited policy relevance in a given region, such as gaming in contexts where basic connectivity is the primary concern, may produce more actionable scores.

\textbf{Urban vs Rural areas}: The urban-rural dimension remains unexplored at scale; comparing threshold sensitivity at the city versus national level would clarify whether score volatility at the national level is partly driven by the mixing of heterogeneous urban and rural connectivity.

\textbf{IQB and external indicators}: Correlating IQB scores with external indicators such as GDP per capita, ITU broadband penetration statistics, and ISP market concentration measures would help establish the concurrent validity of the framework and assess whether its outputs are suitable for use in policy reports and investment decisions.

\bibliographystyle{acm}
\bibliography{reference}

\end{document}